\documentclass[preprint,showpacs,preprintnumbers,amsmath,amssymb]{revtex4}

\usepackage{graphicx}% Include figure files
\usepackage{dcolumn}% Align table columns on decimal point
\usepackage{bm}% bold math

\usepackage[dvips]{color}
\usepackage{ulem}

\begin{document}

%% \preprint{APS/123-QED}

\title{Numerical renormalization group approach to a quartet quantum-dot array connected to reservoirs:\\ 
gate-voltage dependence of the conductance 
}
% Force line breaks with \\

\author{Yunori Nisikawa}
%\email{nisikawa@sci.osaka-cu.ac.jp}
% \altaffiliation[Also at ]{Physics Department, XYZ University.}
\author{Akira Oguri}%
%\email{oguri@sci.osaka-cu.ac.jp}
\affiliation{Department of Material Science, Osaka City University, 
Sumiyoshi-ku, Osaka 558-8585, Japan}%
                                                                                
%\author{A. Hewson}
%\affiliation{

%}%

\date{\today}

\begin{abstract}

The ground-state properties of 
quartet quantum-dot arrays 
are studied using the numerical renormalization group (NRG) method
with a four-site Hubbard model connected to two non-interacting leads. 
Specifically, we calculate the conductance and local charge in the dots 
from the many-body phase shifts, which can be deduced from 
the fixed-point eigenvalues of NRG.
 As a function of
the on-site energy $\epsilon_d$ 
which corresponds to the  gate voltage, 
 the conductance 
shows 
alternatively wide peak and valley. 
Simultaneously, 
 the total number of electrons $N_{\rm el}$
in the four dots 
shows a 
quantized 
stair case behavior 
due to a
large Coulomb interaction $U$.
The conductance 
plateaus of the Unitary limit 
emerging for odd $N_{\rm el}$ 
are caused by the Kondo effect.  
The valleys of the conductance
emerge for even $N_{\rm el}$,
and  their width becomes substantially
large at half-filling. 
It can be regarded as a kind of the Mott-Hubbard insulating behavior 
manifesting in a small system.
These structures of the plateaus and valleys become 
weak for large values of
the hybridization strength 
$\Gamma$ between the chain and leads. 
We also discuss 
the parallel conductance for the array 
connected to four leads.
\end{abstract}


\maketitle

\section{Introduction}
The Kondo effect 
in quantum dots is a subject of current interest, and 
early theoretical predictions \cite{JETP,Ng} 
 have already been confirmed by the experiments.\cite{Gold,Cron}
Recently, the interplay of various effects such as 
Aharanov-Bohm, Fano, Josephson, and Kondo effects have been studied 
extensively, 
\cite{IS0,Hofstetter,Iye,RozhkovArovas,ClerkAmbegaokar,Choi,OTH}
and interesting phenomena 
caused by inter-electron interactions have been expected to be seen at low temperatures.

%%%%%%%%%%%%%%%%%%%%%%%%%%%%%%%%%%%
%method:Theory
For studying the low-temperature transport of 
interacting electron systems, careful calculations are required. 
So far, some numerical approaches to conductance through small 
interacting systems have been examined by several groups.
For instance, a Fermi-liquid based method, \cite{RejecRamsak}
density matrix renormalization method, \cite{Molina,Meden}
and functional renormalization method, \cite{Meden,Meden1}
seem to be complement each other to give reliable information valid at
low temperatures.
%%%Wilson-NRG
 The Wilson numerical renormalization group (NRG) 
method, which successively eliminates higher-energy states 
to obtain accurately the low-lying energy states, 
\cite{Wilson,Kuru1,Kuru2}
has also been applied successfully 
to single and double quantum dots.\cite{IS0,IS1,IS2}
%%%%%%%%%%%%%%%%%%%
%our group:history%
%%%%%%%%%%%%%%%%%%%
%%
Recently, 
we have provided an explicit prescription 
to deduce the many-body phase shifts 
from the fixed-point energy levels of NRG.
 It is applicable to a wide class of 
the quantum-dot systems connected to noninteracting leads. 
With this method, we examined precisely the conductance of
a Hubbard chain of finite size $N_C$ connected to noninteracting 
leads,\cite{OH,ONH}
the low-temperature Fermi-liquid regime of which had been studied 
before with a perturbation theory.\cite{O1,O2}
 It was confirmed that 
at half-filling 
the conductance 
through 
the four dots, $N_C=4$, decreases exponentially with 
increasing $U$ reflecting a Mott-Hubbard type insulating 
behavior.\cite{OH} 
 In contrast to the chains in the even size, for odd $N_C$ 
the Kondo resonance emerges at the Fermi level, and it contributes to 
the Unitary-limit conductance $g =2e^2/h$.

%%%%%%%%%%%%%%%%%%%%%%%%%%
%present cal. conductance%
%%%%%%%%%%%%%%%%%%%%%%%%%%
%%
In quantum dots,
 the on-site potential $\epsilon_d$ 
is a tunable parameter, and it corresponds to the gate voltage.
We have reported the $\epsilon_d$ dependence 
of the conductance of the triple quantum dots.\cite{ONH}
The purpose of the work is 
to study how the conductance through the four quantum-dot chain
behaves as a function of the gate voltage.
Owing to the Fermi-liquid properties 
at low temperatures, the conductance $g$ and local charge $N_{\rm el}$ 
in the quantum dots are determined by the two phase shifts 
$\delta_{\rm even}$ and $\delta_{\rm odd}$, 
which are defined with respect to  the even and odd ($s$ and $p$)
 partial waves, respectively.\cite{ONH}
We calculate these two phase shifts from 
the fixed-point eigenvalues of NRG.
Furthermore, from these phase shifts, we can also deduce 
a parallel conductance through the quantum dots 
connected transversely to four noninteracting leads.

%%%%%%%%%%
%results%%
%%%%%%%%%%
The results of the dc conductance show 
the typical Kondo plateaus of the Unitary
limit $g \simeq 2e^2/h$ when 
the total number of electrons $N_{\rm el}$ in 
the four dots is odd.
On the other hand, the conductance shows wide minima 
for even $N_{\rm el} \simeq 2,\,4,\,6$.
The local charge $N_{\rm el}$ shows a quantized staircase 
behavior as a function of $\epsilon_d$. 
Among the conductance minima, 
the one at half-filling $N_{\rm el} \simeq 4$ 
is the widest, and it can be regarded as a kind of 
the Mott-Hubbard insulating behavior. 
The feature of the conductance near half-filling 
is quite different whether the number of quantum dots is even or odd, 
namely the Kondo plateau appears for odd $N_C$. 

%%%%%%%%%%
%contents%
%%%%%%%%%%
This paper is organized as follows.
In Sec.\ \ref{sec:model},
we describe the outline of the formulation to deduce the conductance 
from the fixed-point energy levels of the NRG.
In Sec.\ \ref{sec:result}, we show the NRG results.
In Sec.\ \ref{sec:discuss}, discussions are given.

\section{\label{sec:model}Model and Formulation}
%%%%%%
%Model
In this section, we describe briefly the relation between the phase shifts
and fixed-point Hamiltonian of NRG.\cite{ONH}
 As far as the zero-temperature value is concerned, 
 the conductance can be calculated more accurately with 
this formulation 
than to calculate it directly from the current-current 
correlation function.

\begin{figure}
\includegraphics[width=0.8 \linewidth]{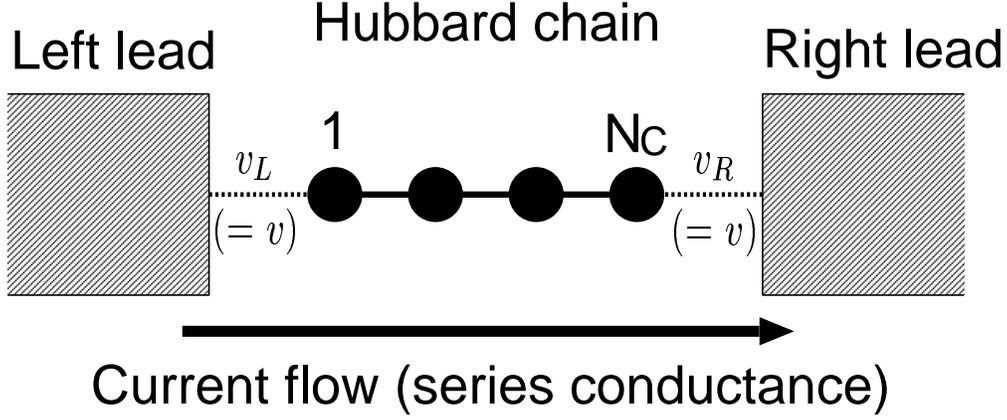}
\caption{\label{fig:system}Schematic picture of a series connection.}
\end{figure}

We start with a Hubbard chain of a finite size $N_C$,  
as illustrated in Fig.\ \ref{fig:system}.
It is connected to two non-interacting leads on the left($L$) and
right($R$) via the tunneling matrix elements
 $v_L^{\phantom{\dagger}}$ and $v_R^{\phantom{\dagger}}$,  
respectively.
%
%Fig:1:start
%Fig:2:end
% 
The Hamiltonian is given by
\begin{align}
{\cal H}   & \, =\,    {\cal H}_{C}^0 \, 
+ \, {\cal H}_{C}^U 
+ \,  {\cal H}_{\rm mix} \, + \, 
{\cal H}_{\rm lead} 
\label{eq:H}
\end{align}
with 
\begin{align}             
 {\cal H}_{C}^0 \ & \, = \,  
 -t 
 \sum_{i=1}^{N_C-1}\sum_{\sigma}  
 \left(\,
 d^{\dagger}_{i\sigma}d^{\phantom{\dagger}}_{i+1\sigma}  
 \,+\,
 d^{\dagger}_{i+1\sigma}d^{\phantom{\dagger}}_{i\sigma}  
\right) \nonumber\\
& \qquad \,+\,
\epsilon_d 
\sum_{i=1}^{N_C}\sum_{\sigma} \, 
 d^{\dagger}_{i\sigma}d^{\phantom{\dagger}}_{i\sigma}  
 \;,
%%  -\sum_{i,j =1}^{N_C}\sum_{\sigma} t_{ij}^{\phantom{0}} \, 
%%  d^{\dagger}_{i\sigma}d^{\phantom{\dagger}}_{j\sigma}  \;,
\label{eq:HC^0}
\\
%% \end{align}
% \\
%% \begin{align}
%%
 {\cal H}_{C}^U \ & \, = \,  
 U\sum_{i=1}^{N_C} 
   d^{\dagger}_{i \uparrow}
   d^{\phantom{\dagger}}_{i \uparrow}
   d^{\dagger}_{i \downarrow}
   d^{\phantom{\dagger}}_{i \downarrow} \;,
\label{eq:HC^U}
\\
%% \end{align}
%\\
%% \begin{align}
 {\cal H}_{\rm mix} \,  & \, =\,
v_{L}^{\phantom 0} 
 \sum_{\sigma}   
 \left(\,  
  d^{\dagger}_{1,\sigma} \psi^{\phantom{\dagger}}_{L \sigma}
\,+\,    
\psi^{\dagger}_{L \sigma}   d^{\phantom{\dagger}}_{1,\sigma} 
\,\right)
  \nonumber \\
& \ \ + v_{R}^{\phantom 0} 
 \sum_{\sigma}   
\left(\, 
\psi^{\dagger}_{R\sigma} d^{\phantom{\dagger}}_{N_C, \sigma} 
      \,+\,   
d^{\dagger}_{N_C, \sigma} \psi^{\phantom{\dagger}}_{R\sigma}
      \,\right)   ,
\label{eq:Hmix}
%\end{align}
%
\\
%\begin{align}
 {\cal H}_{\rm lead}  & \, = \,  
\sum_{\nu=L,R} 
 \sum_{k\sigma} 
  \epsilon_{k \nu}^{\phantom{0}}\,
         c^{\dagger}_{k \nu \sigma} 
         c^{\phantom{\dagger}}_{k \nu \sigma}
\,,
\label{eq:H_lead}
\end{align}
where $d_{i\sigma}$ annihilates an electron with spin 
$\sigma$ at site $i$ in the Hubbard chain,   
which is characterized by the nearest-neighbor hopping 
matrix element $t$, onsite energy $\epsilon_d$, 
and Coulomb interaction $U$.  
In the lead on the left or right 
($\nu = L,\, R$), the operator 
$c_{k \nu \sigma}^{\dagger}$ creates an conduction electron 
with energy $\epsilon_{k\nu}$, the wavefunction for which is denoted by
$\phi_{k\nu} (r)$. 
The linear combinations of the conduction electrons
$\psi_{L \sigma}^{\phantom{\dagger}}$ and 
$\psi_{R \sigma}^{\phantom{\dagger}}$  mixed with the electrons 
 in the interacting sites at $i=1$ and $N_C$, respectively, 
where  $\psi_{\nu \sigma}^{\phantom{\dagger}} 
= \sum_k c_{k \nu \sigma}^{\phantom{\dagger}} 
\, \phi_{k\nu} (r_{\nu})$ and 
$r_{\nu}$ is the position at the interface 
in the lead $\nu$. 
%
%
%% We assume that each of the two leads has a continuous energy spectrum
%% and the density of states $\rho_{\nu}$ ( $\nu= L,\, R$) is a constant 
%% around Fermi level.
%%  The energy scale of the level-broadening
%% becomes $\Gamma_{\nu}= \pi\rho_{\nu} v_{\nu}^2$  ( $\nu= L,\, R$).
We assume that the hybridization strength 
%% the energy scale of the level-broadening 
$\Gamma_{\nu} \equiv \pi v_{\nu}^2 \sum_k 
\left|\phi_{k\nu} (r_{\nu})\right|^2
\delta (\omega - \epsilon_{k \nu}^{\phantom{0}})$ 
is a constant independent of the frequency $\omega$,  
and take the origin of the energy to be $\mu=0$.

\subsection{Ground-state properties and phase shifts }
%inversion
In the following, we assume that the system has an inversion symmetry,
$v_{L} = v_{R}$  ($\equiv v$) 
and $\Gamma_L = \Gamma_R$ ($\equiv \Gamma$). 
Then,  it is convenient to introduce 
the orbitals which have even and odd parities 
\begin{align}
a_{j,\sigma} &\,=
 \frac{
 d_{j,\sigma} \,+\, d_{N_C-j+1,\sigma} }
{\sqrt{2}}
\;,
 \label{eq:orbitals_a}
 \\
b_{j,\sigma} &\,= 
\frac{d_{j,\sigma} \,-\, d_{N_C-j+1,\sigma}}{\sqrt{2}}\;
,
 \label{eq:orbitals_b}
\end{align}
where $j=1,2, \ldots, N_C/2$ for even $N_C$. 
For odd $N_C$, 
%}
there exists one extra 
unpaired orbital $a_{(N_C+1)/2,\sigma} \equiv d_{(N_C+1)/2,\sigma}$ 
in addition to the pairs 
labeled as
 $j=1,2, \ldots, (N_C-1)/2$. 
Due to the inversion symmetry, at zero temperature $T=0$ 
and Fermi energy $\omega=0$, 
the retarded Green's functions for $a_{1,\sigma}$ and $b_{1,\sigma}$
can be written in the forms,\cite{ONH} 
\begin{align}
 \langle\!\langle a_{1,\sigma}^{\phantom{\dagger}}; 
 a_{1,\sigma}^{\dagger} \rangle\!\rangle_{\omega=0}^{\phantom{|}}
&
\, = \, \frac{1}{\Gamma} \, \frac{1}{\kappa_{\rm even} + i } 
\,\equiv \,    
\frac{1}{\Gamma} \, 
\frac{-\,e^{i \delta_{\rm even}}}{\sqrt{\kappa_{\rm even}^2 + 1}}
\;,
\label{eq:G_even}
\\
 \langle\!\langle b_{1,\sigma}^{\phantom{\dagger}}; 
 b_{1,\sigma}^{\dagger} \rangle\!\rangle_{\omega=0}^{\phantom{|}}
&
\, = \, \frac{1}{\Gamma} \, \frac{1}{\kappa_{\rm odd} + i}
 \,\equiv\,
\frac{1}{\Gamma} \, 
\frac{-\,e^{i \delta_{\rm odd}}}{\sqrt{\kappa_{\rm odd}^2 + 1}} 
\;.
\label{eq:G_odd}
\end{align}
Namely, each of these two retarded Green's functions is determined by 
a single parameter, $\kappa_{\rm even}$ or $\kappa_{\rm odd}$, 
which contains all effects of the scatterings and interactions. 
The phase shifts  $\delta_{\rm even}$ and $\delta_{\rm odd}$,
 correspond to the angle of these two Green's functions in the complex plane.
%%
%% (\ref{eq:G_odd}) and (\ref{eq:G_odd})
%%
%
%
%Fig:2:start
\begin{figure}
\includegraphics[width=0.75 \linewidth]{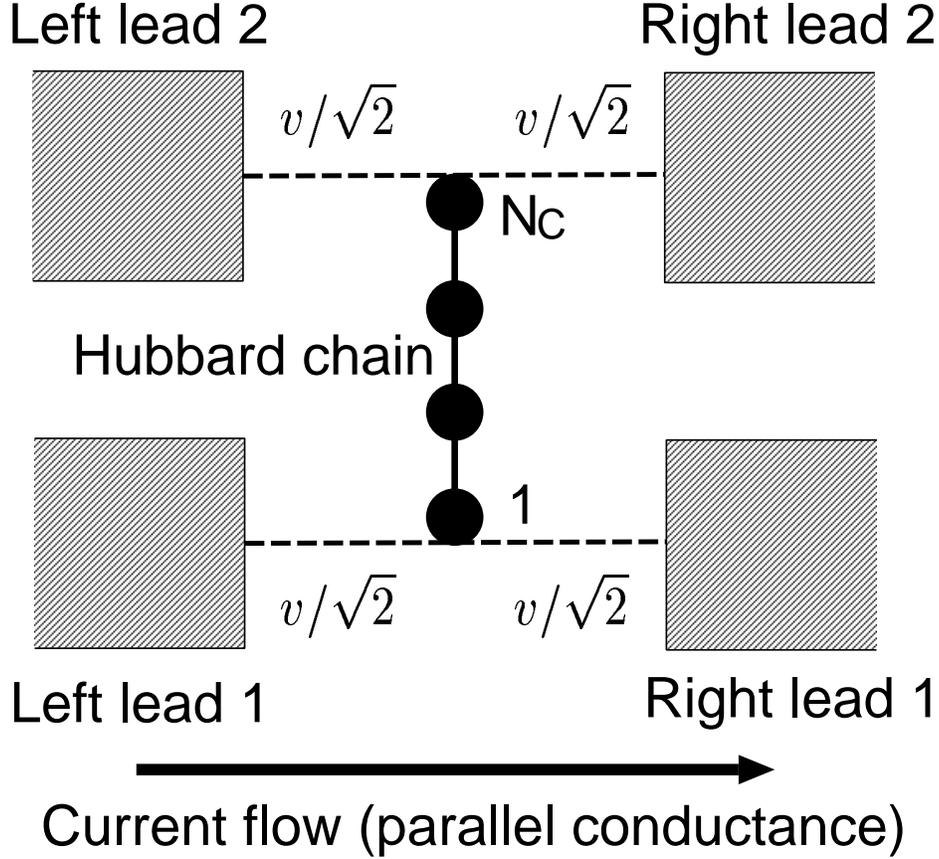}
\caption{\label{fig:parallel}Schematic picture of a parallel connection.}
\end{figure}
%Fig:2:end

%conductance
The two phase shifts  determine the 
ground-state properties of the series connection  
illustrated in Fig.\ \ref{fig:system}.\cite{Kawabata}
Specifically,  using the Kubo formalism and Friedel sum rule,
the dc conductance  $g_{\rm series}$ and 
total number of electrons in the Hubbard chain 
$N_{\rm el}\, \equiv \,\sum_{i=1}^{N_C}\sum_{\sigma} \langle d_{i\sigma}^{\dagger}d_{i\sigma}^{\phantom{\dagger}}
\rangle$ can be expressed in the forms,\cite{ONH}
\begin{align}
g_{\rm series}\, &=\, \frac{2e^{2}}{h}\,
\sin^{2} \Bigl(\delta_{\rm even}
-\delta_{\rm odd}\Bigr)\;,
\label{eq:gs}
\\
N_{\rm el}&
\,=\,\frac{2}{\pi}\Bigl(\delta_{\rm even}+\delta_{\rm odd}\Bigr)\;.
\label{eq:N_el}
\end{align}
These formulas are justified if 
the imaginary of the $N_C \times N_C$ matrix version of the 
retarded self-energy due to the interaction ${\cal H}_{C}^U$ vanishes,
$\mbox{Im}\, \mbox{\boldmath $\Sigma$}^+(0)=0$,  
 at $T=0$ and $\omega=0$.\cite{O2,ONH}

%% 
%% \begin{equation}
%% g_{\rm series}^{\phantom{0}}\, = \, \frac{2 e^2}{h}  
%%     4\Gamma^2 \left| G_{N_C 1}^{+}(0)\right|^2  
%%         \;. 
%% \end{equation}
%%
%% which is defined with respect to the interacting sites
%% $
%% G_{ij}^+(\omega)=
%%  \langle\!\langle d_{i,\sigma}^{\phantom{\dagger}}; 
%%  d_{j,\sigma}^{\dagger} \rangle\!\rangle_{\omega}^{\phantom{|}}
%% $.
%%
%%
% The dc conductance $g_{\rm parallel}$ for the parallel 
% connection as shown in Fig.\ \ref{fig:parallel}, where
% the tunneling matrix elements between the Hubbard chain and
% each lead are given by $v/\sqrt{2}$  for the parallel connection.
% In this case, one can calculate the conductance $g_{\rm parallel}$ and 
% $N_{\rm el}$ for the parallel connection from the phase shifts
% functions obtained for the series connection
%% 
%%
Furthermore, from these two phase shifts defined with respect to 
the series connection, one can deduce 
the parallel conductance $g_{\rm  parallel}$ for the current flowing in 
the horizontal direction in 
the geometry illustrated in Fig.\ \ref{fig:parallel};  
\begin{equation}
g_{\rm  parallel} \,= \,\frac{2e^{2}}{h} \,
 \Bigl(\sin^{2}\delta_{\rm even} \, + \,
 \sin^{2}\delta_{\rm odd}\Bigr)\;.
      \label{eq:gp}
\end{equation}
This expression 
is equivalent to the multi-channel version 
of the Landauer formula,\cite{Landauer}
and it can also
be derived from the Kubo formula. 
\cite{ONH,tanaka1}
Since two conducting channels can contribute to the total current, 
the maximum possible value of  $g_{\rm  parallel}$  is  $4e^2/h$. 
Note that the tunneling matrix elements between the leads and 
interacting part are taken to be $v/\sqrt{2}$ 
as shown  in Fig.\ \ref{fig:parallel}. Because of 
the inversion symmetry between left lead 1 (2) and right lead 1 (2), 
an odd combination of the states from these two leads is separated,
and the interacting site for $i=1$ ($i=N_C$) is coupled 
to an even combination by the matrix element $v$.
Therefore, the ground state properties for the parallel geometry 
are determined by the phase shifts 
$\delta_{\rm even}$ and $\delta_{\rm odd}$ 
that are defined with respect to the series connection in  
eqs.\ (\ref{eq:G_even}) and (\ref{eq:G_odd}). 
Specifically, $N_{\rm el}$ for parallel connection 
is equal to that for the series connection, and is given by eq.\ (\ref{eq:N_el}).
%%
%% 
%% an odd combination of the noninteracting 
%% orbtials can be separated, and 
%% the Green's functions in the central region 
%% $\mbox{\boldmath $G$}^+(\omega)$ coincide with 
%% that for the series connection
%%  
%% Alternatively, 
%% it can be expressed in the form, $g_{\rm parallel}^{\phantom{0}} =
%% (2 e^2 / h) \, 2 \Gamma 
%% \left\{ -\mbox{Im}\, G_{11}^+(0) \right\}$.

\subsection{Fixed-point Hamiltonian and phase shifts}
%NRG
%% 
The  two phase shifts 
$\delta_{\rm even}$ and $\delta_{\rm odd}$ 
can be deduced from 
the fixed-point eigenvalues NRG. In the NRG approach, 
a sequence of the Hamiltonian $H_N$ is introduced,
by carrying out the logarithmic discretization 
for the conduction band with the half-width $D$,\cite{Wilson,Kuru1,Kuru2}
as
\begin{align}
H_N & \,=\,  \Lambda^{(N-1)/2}  
\left( \,
      {\cal H}_{C}^0  + {\cal H}_{C}^U 
 +  H_{\rm mix}^{\phantom{0}}  +   H_{\rm lead}^{(N)}
 \,\right) ,
\label{eq:H_N}
\end{align} 
%\\
\begin{align}
H_{\rm mix}^{\phantom{0}} & \, = \, \bar{v} \, 
       \sum_{\sigma}
\left(\,
f^{\dagger}_{0,L\sigma} d^{\phantom{\dagger}}_{ 1,\sigma}
\,+\, 
d^{\dagger}_{ 1,\sigma}  f^{\phantom{\dagger}}_{0,L\sigma} 
     \right)
     \nonumber \\
 & \quad  +   
       \bar{v}\, 
       \sum_{\sigma} 
       \left(\,
f^{\dagger}_{0,R \sigma} d^{\phantom{\dagger}}_{N_C, \sigma} 
   \,+\, 
d^{\dagger}_{N_C, \sigma} f^{\phantom{\dagger}}_{0,R \sigma} 
         \,  \right)  \;,
\label{eq:H_mix_NRG}
\\
H_{\rm lead}^{(N)} &\,=\,
D\,{1+1/\Lambda \over 2} \,
\sum_{\nu=L,R}
\sum_{\sigma}
\sum_{n=0}^{N-1} 
\, \xi_n\, \Lambda^{-n/2}
\nonumber \\
& \qquad \quad \times  
\left(\,
  f^{\dagger}_{n+1,\nu\sigma}\,f^{\phantom{\dagger}}_{n,\nu\sigma}
  +  
 f^{\dagger}_{n,\nu\sigma}\, f^{\phantom{\dagger}}_{n+1,\nu\sigma}
 \,\right) \;,
\label{eq:H_lead_NRG}
\end{align}
where $\,\bar{v}=\sqrt{ 2D \Gamma A_{\Lambda}/\pi }\,$,
$\, A_{\Lambda} = \frac{1}{2} 
 {1+1/\Lambda \over 1-1/\Lambda }
\,\log \Lambda\,$, and 
\begin{equation}
\xi_n \,=\, 
\frac{1-1/\Lambda^{n+1}}
{ \sqrt{1-1/\Lambda^{2n+1}}\sqrt{1-1/\Lambda^{2n+3}} } \;.
\end{equation}
% $
% \xi_n = (1-1/\Lambda^{n+1})
% /(\sqrt{1-1/\Lambda^{2n+1}}\sqrt{1-1/\Lambda^{2n+3}})$.
%%
%% For the finite Hubbard chain we are considering, 
%% the low-lying energies of $H_N$ converge for large $N$ 
%% to the spectrum that can be described by 
%% a local Fermi liquid \cite{OH,ONH}, 
%% and the fixed-point eigenvalues can be reproduced by 
%% an effective Hamiltonian for free quasi-particles
%%
As in the case of the single Anderson (or Kondo) impurity, 
\cite{Wilson,Kuru1,Kuru2} 
the low-lying eigenvalues of $H_N$ for 
the finite Hubbard chain converge, 
for large $N$, to the fixed-point values which have 
one-to-one correspondence to the free quasi-particles 
of a local Fermi liquid.\cite{OH,ONH}   
The fixed-point Hamiltonian describing 
the free quasi-particles can be written in the form
\begin{equation}
H_{\rm qp}^{(N)}   \, = \,  
\Lambda^{(N-1)/2}  
\left(
H_{C}^{\rm eff} + H_{\rm mix} + H_{\rm lead}^{(N)} 
\right) \;,
\label{eq:Hqp}
\end{equation}
 where 
 $H_{C}^{\rm eff} \equiv   {\cal H}_{C}^0 
+ \sum_{ij=1}^{N_C} 
\mbox{Re}\, \Sigma^{+}_{ij}(0)\, 
d^{\dagger}_{i \sigma} d^{\phantom{\dagger}}_{j \sigma}$.
The many-body corrections enter through the real part of 
self-energy at $T=0$, $\omega=0$. 
%% 
%% , which is defined with respect to the interacting sites
%%
%% It enables us to deduce the parameter $\kappa_{\gamma}$ or 
%% phase shifts $\delta_{\gamma}$.
%%
%% The Hamiltonian $H_{\rm qp}^{(N)}$  describes 
%% the free quasi-particles in the system 
%% consisting of $N_C + 2(N+1)$ sites.
%%
%% The corresponding Green's function 
%% $\mbox{\boldmath $G$}_{\rm qp}(\omega)$ for the interacting sites 
%% determines 
%% the eigenvalue $\varepsilon^*$ of $H_{\rm qp}^{(N)}$ 
%% by the condition 
%% $\det \left\{\mbox{\boldmath $G$}_{\rm qp}
%% (\varepsilon^*)\right\}^{-1} = 0$. 
%
%% Owing to the inversion symmetry, this equation can be factorized 
%% and the eigenstates of $H_{\rm qp}^{(N)}$ can be classified 
%% according to the parity.
%%
%{\color{blue}\sout{
%However, 
%instead of calculating the self-energy directly, 
%}}
With NRG, one can calculate accurately 
the quasi-particle energies $\varepsilon^*_{\gamma}$ 
with parity $\gamma$ (= \lq\lq even" or \lq\lq odd")   
from the fixed-point eigenvalues of $H_N$. 
Then, the parameters $\kappa_{\rm even}$ and $\kappa_{\rm odd}$ 
defined in eqs.\ (\ref{eq:G_even}) and (\ref{eq:G_odd}) 
can be deduced from the quasi-particle 
energy $\varepsilon^*_{\gamma}$,\cite{ONH}
\begin{equation}
\kappa_{\gamma}\,=\,\left(\bar{v}^2/\Gamma D\right) \lim_{N\to
 \infty}D\Lambda^{(N-1)/2}\,
 \mbox{\sl g}_{N}^{\phantom{|}}(\varepsilon^*_{\gamma})\;.
\label{eq:kappa2}
\end{equation}
Here, $\mbox{\sl g}_N^{\phantom{|}} (\omega)
=\sum_{m=0}^{N} |\varphi_m(0)|^2/(\omega- \epsilon_m)$ is
the Green's function for 
one of the isolated leads
described in $H_{\rm lead}^{(N)}$. 
The eigenvalue and eigenfunction
of for one isolated lead are denoted by 
 $\epsilon_m$ and $\varphi_m(n)$, respectively, for $0\leq n \leq N$.

\section{\label{sec:result} NRG results for the ground-state properties}

\begin{figure}
\includegraphics[width=8cm,height=6cm]{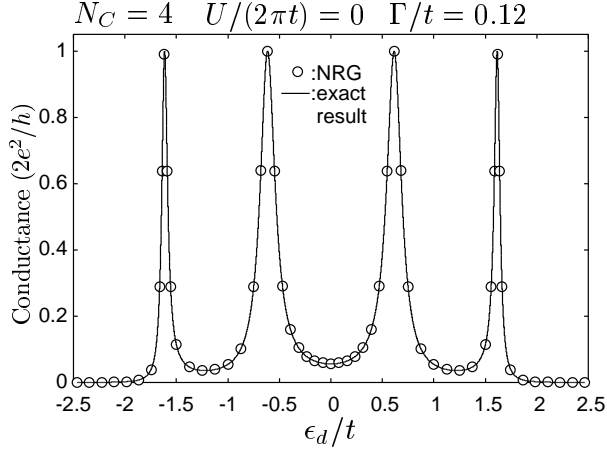}
\caption{\label{fig:U0} 
The conductance $g_{\rm series}$ calculated
 using NRG method (circle dots) compared with the exact result(solid line).
 Here, $U/(2\pi t)=0$ and $\Gamma/t=0.12$. For NRG, we use 
 $\Lambda=6.0$ and $t/D=0.1$.}
\end{figure}

%
%% In the following, we assume that the bare hopping matrix elements $t_{ij}$,
%% which are defined in eq.\ (\ref{eq:HC^0}) for ${\cal H}_C^0$,
%% are described by the nearest-neighbor hopping $t$
%% and onsite energy $\epsilon_d$ as 
%% \begin{align}             
%%  {\cal H}_{C}^0  \,& = \,  
%%  -t 
%%  \sum_{i=1}^{N_C-1}\sum_{\sigma}  
%%  \left(\,
%% d^{\dagger}_{i\sigma}d^{\phantom{\dagger}}_{i+1\sigma}  
%% \,+\,
%%  d^{\dagger}_{i+1\sigma}d^{\phantom{\dagger}}_{i\sigma}  
%% \right) \nonumber\\
%% & \qquad \,+\,
%% \epsilon_d 
%% \sum_{i=1}^{N_C}\sum_{\sigma} \, 
%%  d^{\dagger}_{i\sigma}d^{\phantom{\dagger}}_{i\sigma}  
%%  \;,
%% \label{eq:HC^0_hubbard}
%% \end{align}
%%
%% In order to classify the eigenstates according to the parity, 
%%
%% by introducing the bonding and anti-bonding orbitals also for the leads
%% $(f^{\dagger}_{n,R\sigma} \pm  f^{\dagger}_{n,L\sigma})/\sqrt{2}$ 
%% in order to 
%
%% When we construct the Hilbert space for the next NRG step,
%% we add the bonding orbitals first and retain all
%% eigenstates after carrying out the diagonalization.
%% Then, we add the remaining anti-boding orbitals,
%
In this section,
we apply the formulation described in the above 
to the four-site Hubbard chain, $N_{C}=4$, attached to the reservoirs.
In each step of the iteration, the Hilbert space for $H_{N+1}$ is
constructed from that for $H_N$ and extra $16$ states with respect to 
the two orbitals 
  $f^{\dagger}_{N+1,R\sigma}$ and $f^{\dagger}_{N+1,L\sigma}$. 
We have kept the lowest 1716 eigenstates for constructing the 
Hilbert space for the next step. 
With this procedure the discretized Hamiltonian $H_N$, which consists of 
$2(N+1)+N_C$ sites, can be diagonalized exactly up to $N=0$ for $N_C
=4$. 
We have checked that 
the conductance of the the noninteracting case $U = 0$
are reproduced sufficiently well with this procedure 
also for a rather large value of 
the discretization parameter $\Lambda =6.0$, 
%%new:start
 as shown in Fig.\ \ref{fig:U0}.
%%new:end
%QP-confirm
We have also confirmed that the numerical 
results for the fixed-point eigenvalues of $H_N$ can be mapped onto 
the energy spectrum of the free quasi-particles in all parameter sets 
we have examined. It justifies the assumption of the local Fermi liquid
 we have made in the previous section.

%% 
%Fig::start
%Fig::end

%Fig::start%%U1.0G0.12H0.1%%
\begin{figure}
\includegraphics[width=8cm,height=12cm]{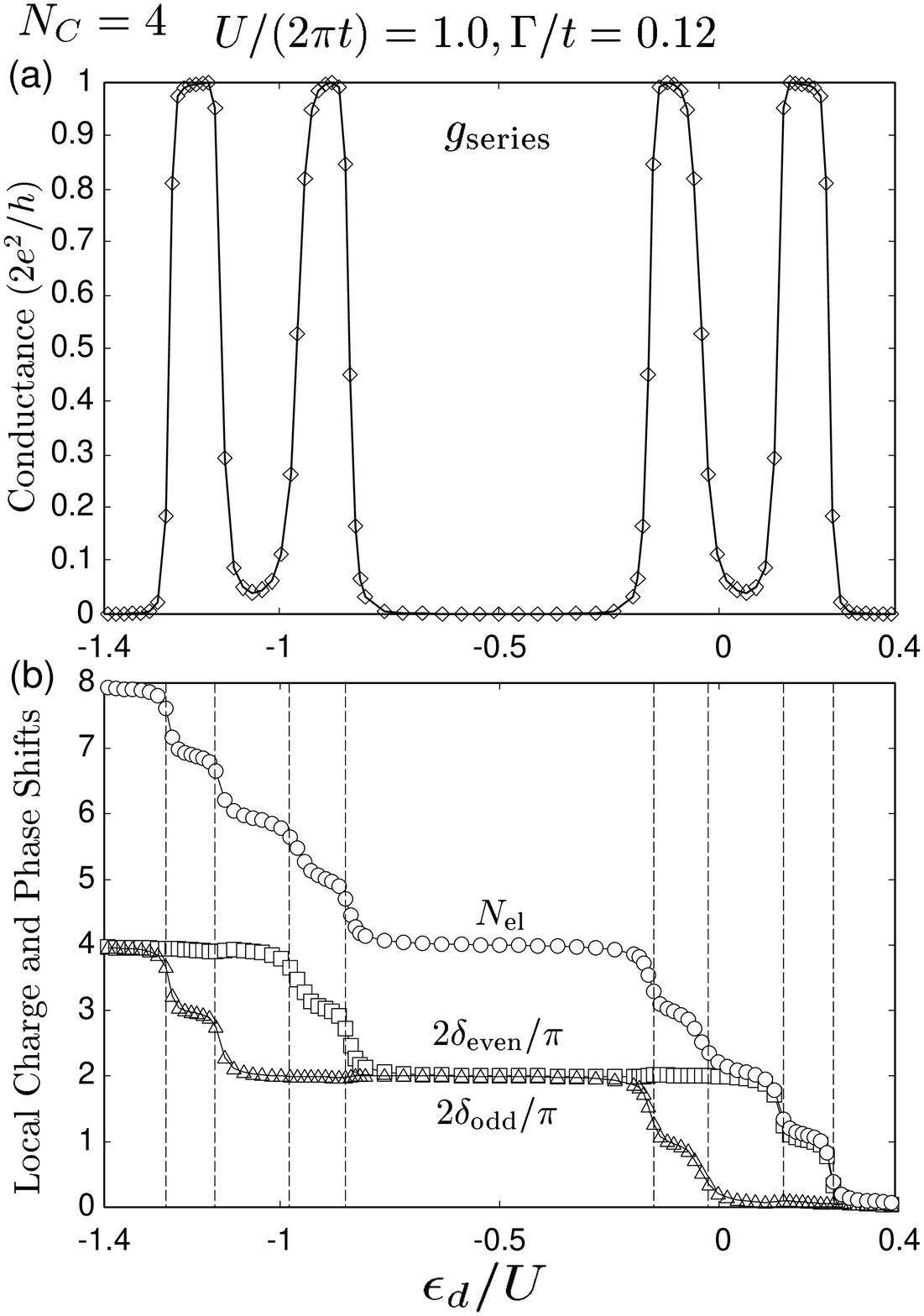}
\caption{\label{fig:U1.0G0.12H0.1}(a) The conductance $g_{\rm series}$,
 (b) 
local charge $N_{\rm el}$, phase shifts $2\delta_{\rm even}/\pi$ and  $2\delta_{\rm odd}/\pi$ 
 as functions of $\epsilon_d/U$ 
for $U/(2\pi t)=1.0$, $\Gamma/t = 0.12$, $t/D=0.1$, and $\Lambda=6.0$. 
The dashed vertical lines in (b) 
correspond to the values of $\epsilon_d$, at which $N_{\rm el}$ jumps in the 
limit of $\Gamma \to 0$. 
}
\end{figure}
%Fig::end%%U1.0G0.12H0.1%%
%

%
%Fig::start%%U0.2G0.12H0.1%%
\begin{figure}
\includegraphics[width=8cm,height=12cm]{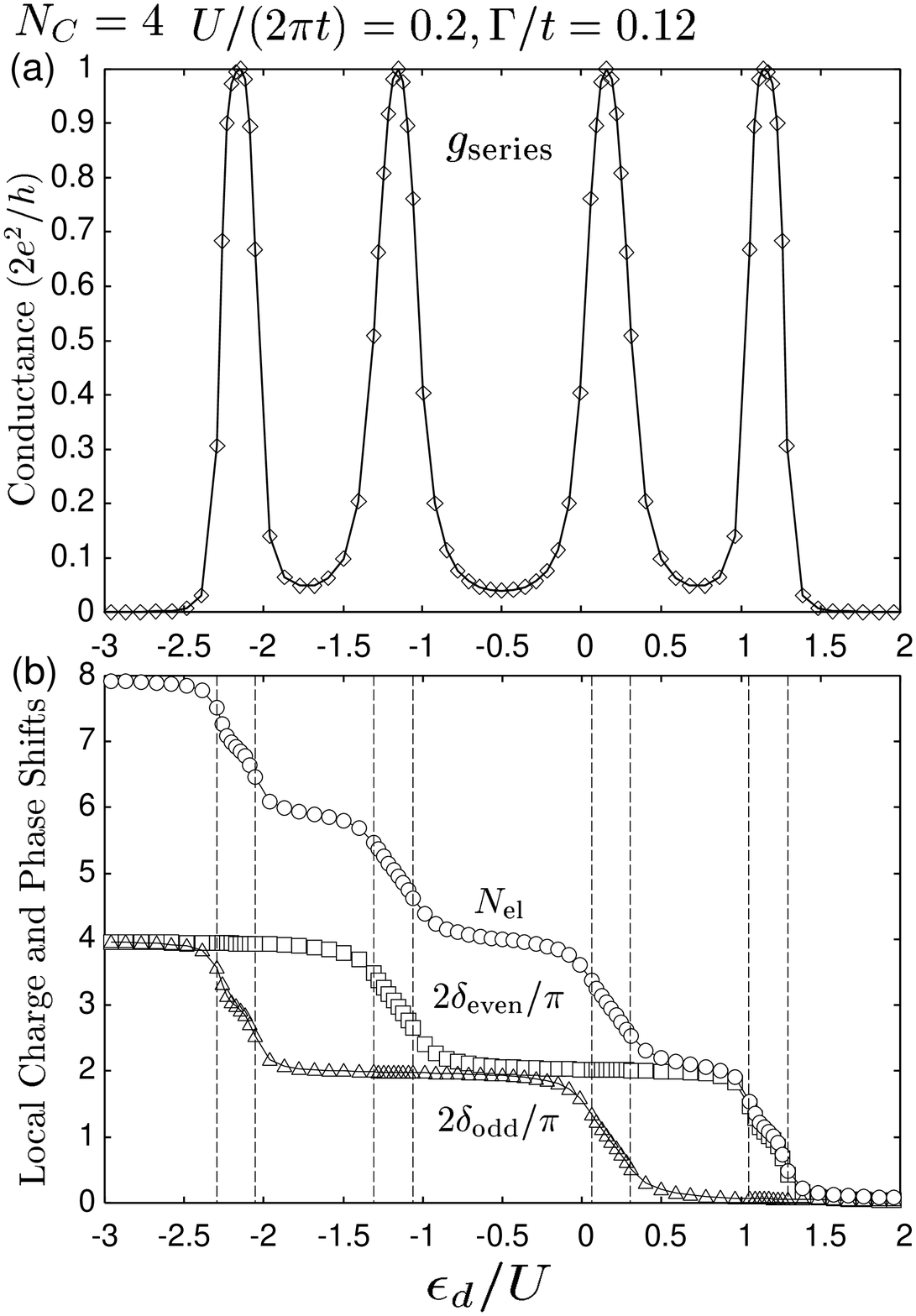}
\caption{\label{fig:U0.2G0.12H0.1}(a) The conductance $g_{\rm series}$, 
(b) local charge $N_{\rm el}$,  phase shifts $2\delta_{\rm even}/\pi$ and  $2\delta_{\rm odd}/\pi$
  as functions of $\epsilon_d/U$. Here, 
$U/(2\pi t)=0.2$, $\Gamma/t = 0.12$, $t/D=0.1$, and $\Lambda=6.0$. 
The dashed vertical lines in (b) correspond to the values of $\epsilon_d$
at which $N_{\rm el}$ jumps in the limit of 
$\Gamma \to 0$.}
\end{figure}
%Fig::end%%U0.2G0.12H0.1%%

\subsection{Series connection}

%%%%%%%%%%%%%%%%%%
%
% Series
%
%%%%%%%%%%%%%%%%%%
%%%U1.0G0.12H0.1%%%
%intro:start
We now show the $\epsilon_{d}$ dependence of 
the conductance of the quantum dots in the series connection. 
As already mentioned, the onsite potential $\epsilon_{d}$ 
is a tunable parameter that is controlled 
by the gate voltage in real quantum dots systems.
%
%Here, we present the calculated results of the conductance 
%for the series connection. 
%intro:end
%
%exp. of Fig.:start
In Fig.\ \ref{fig:U1.0G0.12H0.1}, 
(a) the conductance $g_{\rm series}^{\phantom{0}}$, 
(b) total charge $N_{\rm el}$ in the four-site Hubbard chain, the 
phase shifts $\delta_{\rm even}$ and $\delta_{\rm odd}$ 
are plotted as functions of $\epsilon_d/U$, where the parameters are
chosen to be 
$U/(2\pi t) = 1.0$, $\Gamma/t = 0.12$, $t/D = 0.1$ and $\Lambda=6.0$.  
The vertical dashed lines in (b)  
correspond to the values of $\epsilon_d$ 
at which $N_{\rm el}$ jumps discontinuously 
in the limit of $\Gamma \to 0$. 
%exp. of Fig.:end
%conductance:Kondo
The Kondo plateaus of the Unitary limit 
 $g_{\rm series}^{\phantom{0}} \simeq 2e^2/h$ are seen 
in the regions  corresponding to the odd occupations 
$N_{\rm el} \simeq 1$, $3$, $5$, $7$, where the value of the phase
 shifts is almost locked such that $\delta_{\rm even} -\delta_{\rm odd} \simeq \pi/2$. 
 On the other hand, for even occupancies $N_{\rm el}\simeq 2$, $4$, $6$,
 the conductance is suppressed substantially, and the phase shifts are
 locked at $\delta_{\rm even} -\delta_{\rm odd} \simeq 0$ or $\pi$. 
% 
%Mott-Hubbard
The conductance valley at half-filling becomes wider and deeper than the
other valleys. Although the size of chain is not so large, i.e.,  
$N_C=4$, the feature of the valley at half-filling can be 
understood as a kind of the Mott-Hubbard insulating behavior.
%
%furi
We will 
discuss later the behavior of the conductance near half-filling 
more quantitatively. 
In Fig.\ \ref{fig:U1.0G0.12H0.1}, the hybridization energy scale
$\Gamma$ is 
chosen to be much smaller than the hopping matrix element $t$. 
Therefore, the local charge $N_{\rm el}$ shows 
a clear stair case step behavior. 
In the limit of $\Gamma \to 0$,  
the local charge jumps discontinuously 
at the values of $\epsilon_d$ corresponding to the dashed lines. 
Note that
unless the Coulomb interaction, 
the phase shifts do not show the $\pi/2$ steps 
corresponding to the plateaus for odd $N_{\rm el} =1,3,5,$ and $7$.
%e-o-e-o structure
When the onsite potential $\epsilon_{d}/U$ decreases from $0.4$ to $0$, 
the phase shift for the even partial wave $\delta_{\rm even}$ 
increases from $0$ to $\pi$ via two successive $\pi/2$ steps, 
while that for the odd partial 
wave $\delta_{\rm odd}$ remains almost unchanged at $0$.
Then, in the region of $-0.5< \epsilon_{d}/U < 0.0$, $\delta_{\rm odd}$ shows one
$\pi/2$ step, while the even part remains 
to be $\delta_{\rm even} \simeq \pi$. 
Similar features are seen also for $\epsilon_{d}/U < -0.5$.
In fact, the two regions
$\epsilon_{d}/U < -0.5$ 
and 
$\epsilon_{d}/U > -0.5$ are relating each other via an electron-hole transformation.
In the case of the linear chain of the quantum dots, the resonance
states of the even parity for the even partial wave and that of the odd
one cross the Fermi level alternatively. 
Thus, the $\pi/2$ steps took place 
for even and odd phase shifts 
alternatively.

%%%%%%%%%%%%%%%%%%%%%
%
%Weak U
%
%%%U0.2G0.12H0.1%%
%exp. of Fig.:start
We next examine the ground-state properties for small $U$. 
The results of the conductance and local charge for $U/(2\pi t)=0.2$ 
are shown in Fig.\ \ref{fig:U0.2G0.12H0.1}, where $\Gamma/t = 0.12$ and $t/D = 0.1$. 
%exp. of Fig.:end
%
The peaks of the conductance becomes very narrow 
compared to those for larger $U$ 
shown in Fig.\ \ref{fig:U1.0G0.12H0.1}. Nevertheless, the shape of the
peaks seen deviates from a simple Lorentzian shape as a result of the
many-body effects 
due to the small but finite $U$.
%because $U$ is small.
%
The conductance valley at half-filling $N_{\rm el} \simeq 4$ is 
slightly wider than the other ones at  $N_{\rm el} \simeq 2$ and $6$.
However, the difference is not so large.
In Fig.\ \ref{fig:U0.2G0.12H0.1} (b),
the  steps for the local charge are seen clearly for even occupancies 
$N_{\rm el} \simeq 2, 4,$ and $6$. 
However, for odd occupancies, only a weak structure can be recognized just for 
 $N_{\rm el} \simeq 1$ and $7$. Therefore, 
 when a resonance state passes through the Fermi level,
two electrons occupy the state almost simultaneously. Correspondingly, 
the phase shifts $\delta_{\rm even}$ and $\delta_{\rm odd}$ 
do not show clear $\pi/2$ steps, although the weak structure can be seen for 
$N_{\rm el} \simeq 1$ and $7$.
%%%%

%
%Fig::start
\begin{figure}
\includegraphics[width=8cm,height=6cm]{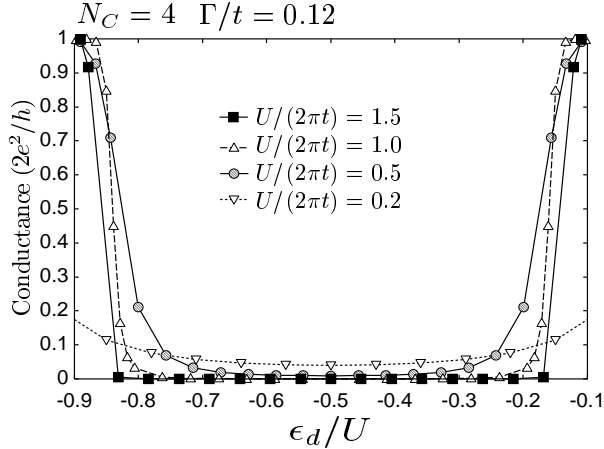}
\caption{\label{fig:U-dep}The 
conductance near half-filling for several values of the Coulomb interaction 
$U/(2\pi t) = 0.2, 0.5, 1.0$ and $1.5$. 
Here,
  $\Gamma/t = 0.12$, $t/D = 0.1$, and $\Lambda=6.0$. }
\end{figure}
%Fig::end

%%%%%%%%%%%%%%%%%%
%
% Mott-Hubbard
%
%%U-dep
%intro:start

%intro:end
%
%exp. of Fig.:start
In order to see how the conductance valley at half-filling 
evolves with the repulsive interaction, the conductance is plotted for several values 
of $U/(2\pi t) = 0.2, 0.5, 1.0$ and $1.5$ in Fig.\ \ref{fig:U-dep},
where the other parameters are chosen to be 
 $\Gamma/t=0.12$, $t/D =0.1$, and $\Lambda =6.0$.
%exp. of Fig.:end
%
%% From Fig.\ \ref{fig:U-dep}, we can find the following facts.
%
%
As $U$ increases, the valley 
becomes wider, and the edge at the both side
becomes sharper. Furthermore, the minimum value of the conductance 
at the electron-hole symmetric case $\epsilon_{d}= -U/2$ 
decreases exponentially with increasing $U$, 
which has been confirmed precisely in a previous work.\cite{OH}
These results imply that for even $N_C$ 
the conductance valley at half-filling approaches  
to the insulating gap, even though the size of the chain examined here
is just $N_C=4$. However, because $N_C$ is finite,
the low-lying energy states of the whole system 
including the noninteracting leads are described by 
the Fermi-liquid fixed point. Therefore, 
the ground state is not an exact insulating state, but a highly
renormalized metallic state with heavy quasi-particles.
%%%%%%%%%%%
%
%%%%%%%%%
%Large N_C
For large even $N_{C}$, however, 
the conductance is expected to show 
an exponential dependence 
$g\propto e^{-N_{C}/\xi}$, 
where $\xi\sim \hbar v_{F}/\Delta_{\rm
gap}$ is a correlation length determined by the Hubbard gap
$\Delta_{\rm gap}$ and Fermi velocity $v_{F}$.
%

\begin{figure}
\includegraphics[width=7.8cm,height=11cm]{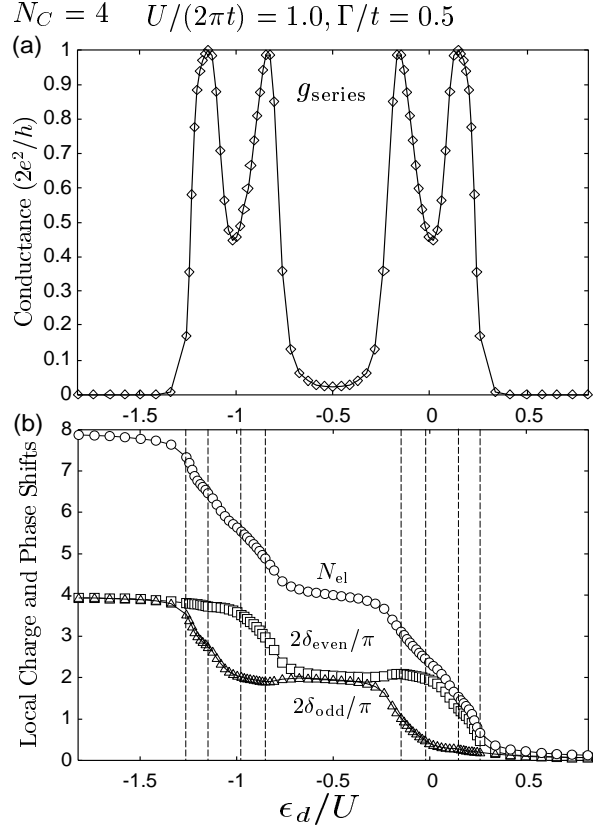}
\caption{\label{fig:U1.0G0.5H0.02}(a) 
The conductance $g_{\rm series}$, (b)local charge $N_{\rm el}$, 
phase shifts $2\delta_{\rm even}/\pi$ and  $2\delta_{\rm odd}/\pi$ 
  as functions of $\epsilon_d/U$. Here, 
$U/(2\pi t)=1.0$, $\Gamma/t = 0.5$, $t/D=0.02$,  and $\Lambda=6.0$. 
The dashed vertical lines in (b) correspond to the values of $\epsilon_d$ 
at which $N_{\rm el}$ jumps in the limit of  $\Gamma \to 0$.}
\end{figure}

%%%%%%%%%%%%%%%%%%
%
% large G
%
%%%U1.0G0.5H0.02%%
%intro:start
%% We also calculate the conductance with varying 
%% the coupling energy between the leads and the Hubbard chain. 
%
%% Here, we consider the case where 
%% the four-site Hubbard chain is coupled to the leads with 
%% larger hybridization than the cases of the above calculations.
%
So far,  
the hybridization energy scale $\Gamma$ 
is taken to be 
much smaller than the hopping matrix element $t$ between 
the dots, $\Gamma= 0.12 t$.
In order to study the ground-state properties for larger $\Gamma$,
 we have carried out the calculations taking the hybridization to be $\Gamma/t = 0.5$.
%intro:end
%
%exp. of Fig.:start
The results are show in Fig.\ \ref{fig:U1.0G0.5H0.02}, where 
$U/(2\pi t)=1.0$,  $t/D =0.02$  and $\Lambda=6.0$.
%exp. of Fig.:end
%
The local charge $N_{\rm el}$, 
which shows a staircase behavior for small $\Gamma$,
 becomes a gentle slope in Fig.\ \ref{fig:U1.0G0.5H0.02} (b).
This is because the large hybridization makes 
the resonance peaks broad,  
and it reduces effectively the correlation effects.
Therefore, the peaks of conductance become round. 
Furthermore, the conductance valleys especially the ones corresponding to  
 $N_{\rm el} \simeq 2$ and $6$ become shallow.
%%
%% it causes screening of $U$ due to conduction electrons.
%U no gap ha tigau --U>G
%
The valley at half-filling is still deeper than 
the other two valleys because of the correlations   
that lead the Mott-Hubbard behavior in the thermodynamic limit.
%%
%%  because the value of the Coulomb interaction $U$ is 
%% larger than that of the hybridization $\Gamma$.
%

%
%Fig::start%%%U1.0G0.5H0.02%%
%Fig::end%%%U1.0G0.5H0.02%%

%Fig::start
\begin{figure}
\includegraphics[width=8cm,height=12cm]{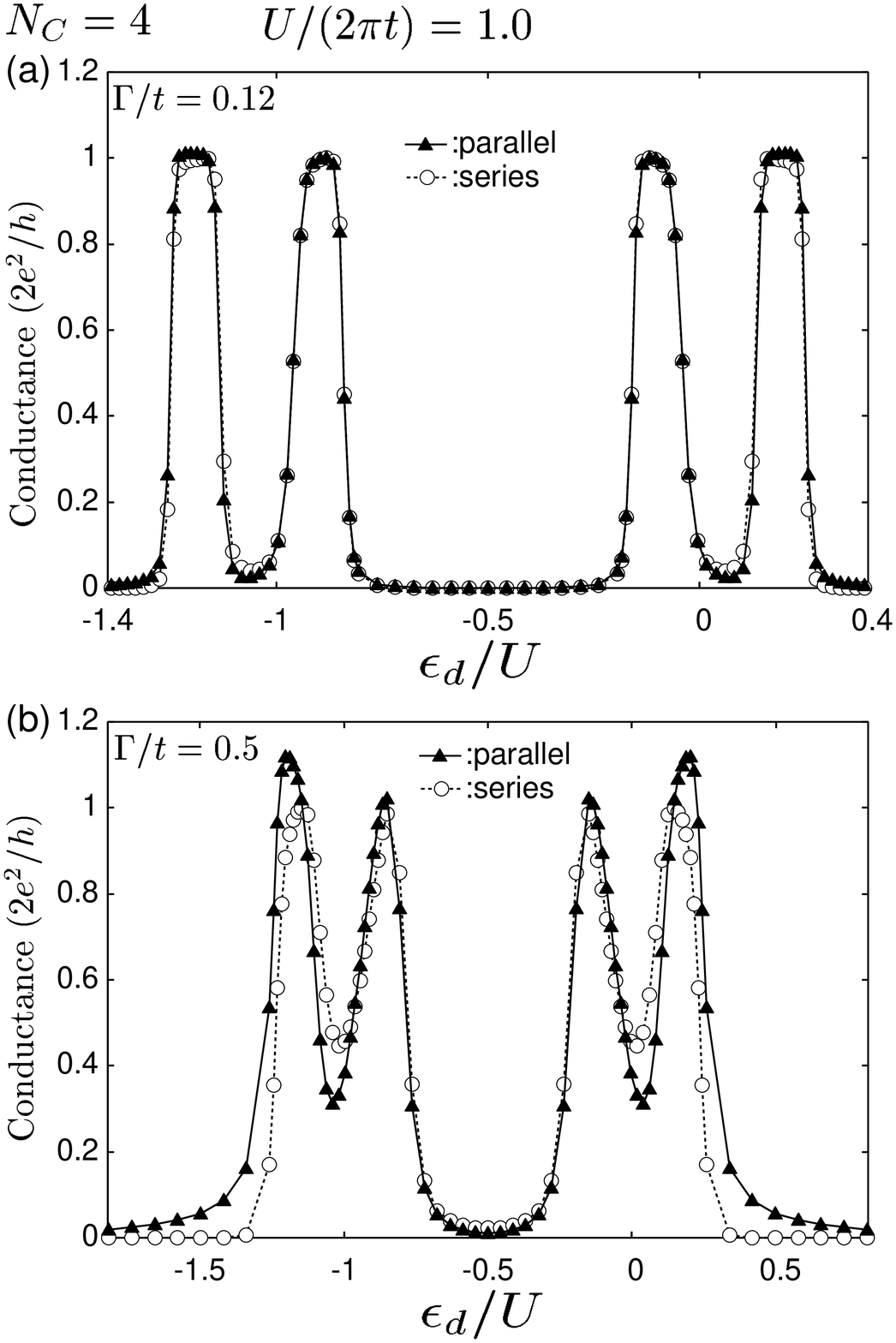}
\caption{\label{fig:para_gam} The  series (solid line)
 and parallel (dotted line) conductances for 
$U/(2 \pi t)=1.0$ 
as functions of $\epsilon_d/U$. Here,
(a) $\Gamma/t=0.12$, $t/D=0.1$, and  (b) $\Gamma/t=0.5$, $t/D=0.02$.}
\end{figure}
%Fig::end
%

\subsection{Parallel connection}

%%%%%%%%%%%%%%%%%%
%
% Parallel
%
%intro:start
%% In this subsection, we present the calculated results of the 
%% conductance for the parallel connection
%%  compared with that of the conductance for the series connection.
%%
We have also calculated the parallel conductance $g_{\rm  parallel}$ 
using eq.\ (\ref{eq:gp}) with the phase shifts obtained for the series connection.
%exp. of Fig.:start
The results are shown in Fig.\ \ref{fig:para_gam} as functions of $\epsilon_d/U$
for $U/(2\pi t)=1.0$. 
In the figures, both the series and parallel conductances 
are plotted for comparison, and for the hybridization
two different values  are examined, (a) $\Gamma/t = 0.12$ and (b) $\Gamma/t = 0.5$.
%exp. of Fig.:end
%%
For small $\Gamma$, the parallel and series conductances are almost coincide with 
each other as seen in Fig.\ \ref{fig:para_gam} (a). 
The difference becomes larger, however, 
for large $\Gamma$, as seen in the lower panel (b). 
The valleys of the conductance are 
deeper for the parallel connection than 
that of the series connection.
This seems to be caused by the fact 
the even and odd parts of the partial waves 
contribute to the parallel conduction  
separately through eq.\ (\ref{eq:gp}).
%% However, the difference is small for small $\Gamma$.
%
%e-o-e-o
The parallel conductance for large $\Gamma$,
 shown in Fig.\ \ref{fig:para_gam} (b), is slightly  
larger than $2e^2/h$ at the first and fourth peaks. 
As mentioned in the previous section,
 the upper bound of $g_{\rm  parallel}$ is $2e^2/h\times 2$,
because two conducting channels can contribute to the current 
in the geometry shown in Fig.\ \ref{fig:parallel}.
However,
in the present case, the parallel conductance does not reach the upper
bound. Similar feature has also been confirmed to be 
seen for the triple dots in a previous work.\cite{ONH}  
As mentioned,
in the case of the linear chain of the quantum dots, 
the resonance state of the even parity
and that of the odd parity passes through the Fermi level alternatively 
when $\epsilon_d$ increases (or decreases). 
Therefore, only a single resonance state can contribute to the tunneling
current for given $\epsilon_d$, so that the peak value of the conductance
for the linear chain is bounded at $g_{\rm  parallel} \simeq 2e^2/h$ 
despite having the two conducting channels.
To reach the Unitary-limit value 
$4e^2/h$ for two conducting channels, the quantum dots are required to  
have a degeneracy in the discrete energy levels. 
It is not the case of the linear chain considered in the present work.

%Fig::start
\begin{figure}
\includegraphics[width= \linewidth]{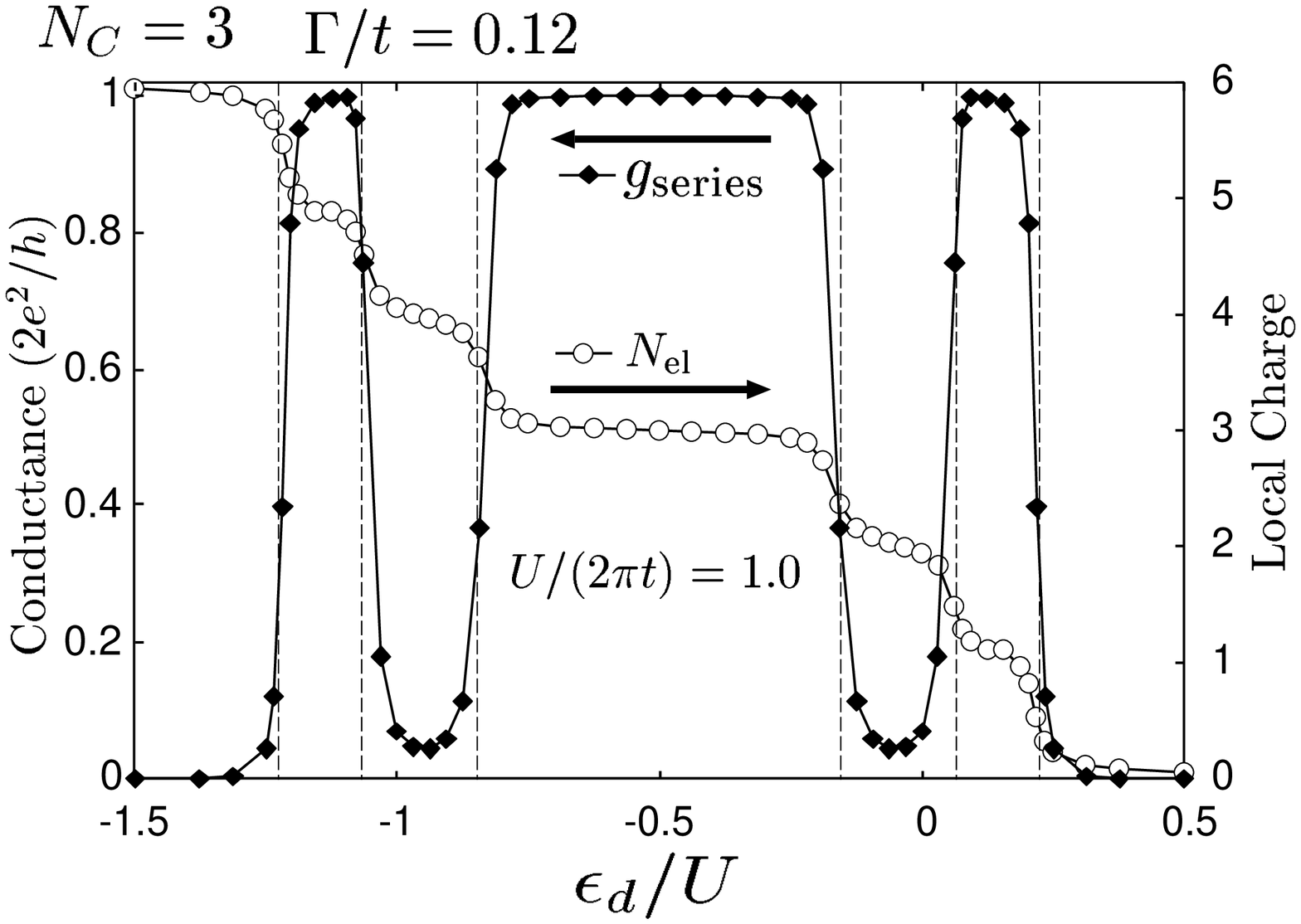}
\caption{\label{fig:3dot} 
 NRG results for the conductance $g_{\rm
 series}$ and local charge $N_{\rm el}$  for triple dots $N_{C}=3$ 
 as functions of $\epsilon_d/U$. Here,
 $U/(2\pi t) = 1.0$,  $\Gamma/t = 0.12$, $t/D = 0.1$, and $\Lambda=6.0$.
The dashed vertical lines correspond to the values of $\epsilon_d$
at which $N_{\rm el}$ jumps in the limit of $\Gamma \to 0$.
%% The lower panel (b) shows the conductance around 
%%  half-filling for $U/(2\pi t) = 0.2$ and $1.0$.
}
\end{figure}
%Fig::end

%Fig::start
\begin{figure}
\includegraphics[width= \linewidth]{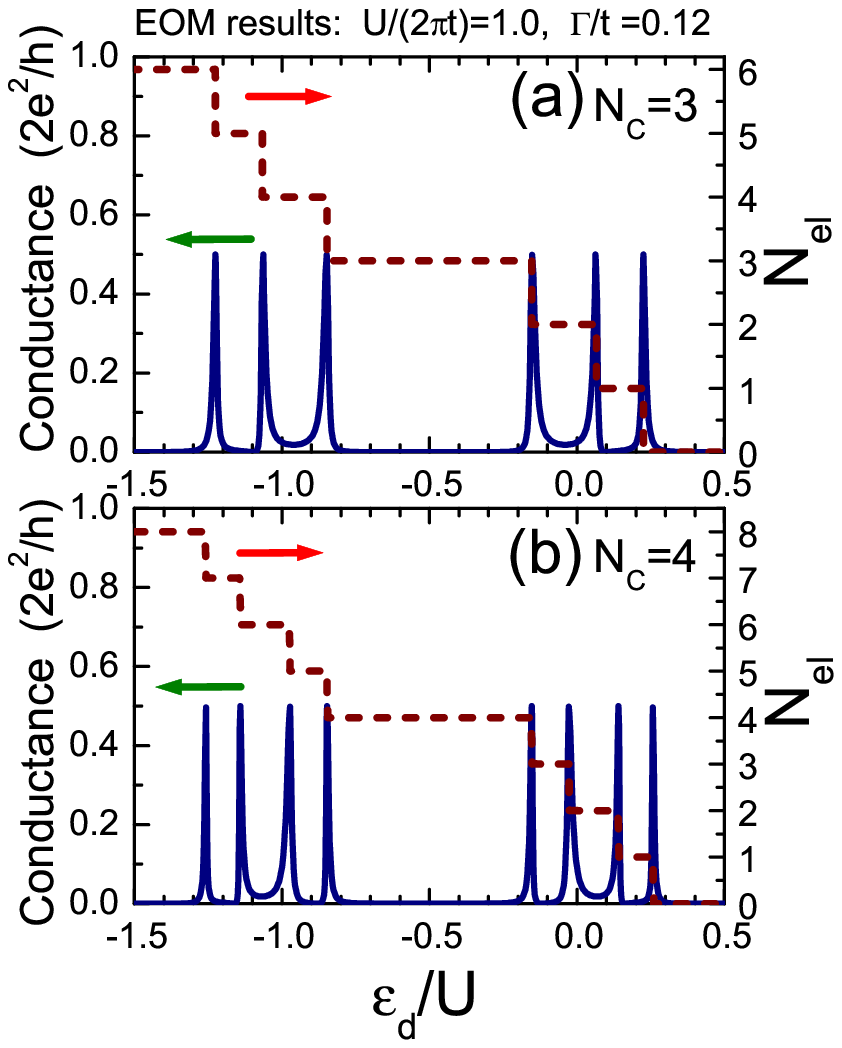}
\caption{\label{fig:EOM}(Color online) 
EOM (equation of motion) results for the series conductance for 
the quantum-dot array of the size 
(a) $N_{C}=3$ and (b) $N_{C}=4$ obtained from eq.\ (\ref{eq:EOM_g}),
where  $U/(2\pi t) = 1.0$ and  $\Gamma/t = 0.12$.
The results describe a typical feature of a Coulomb oscillation 
 at high temperatures $T_K \ll  T \ll U$. 
The dashed lines show the local charge $N_{\rm el}$ 
in the isolated {\it molecule\/} for $\Gamma=0$.
}
\end{figure}
%Fig::end

\section{\label{sec:discuss}Discussions}

%Hubbard Gap through condactace
%Nc :finite odd-->Kondo
%Fig::end
As discussed in Sec.\ \ref{sec:result},
 a kind of the Mott-Hubbard insulating behavior is seen at
half-filling 
in the conductance through the one-dimensional array consisting of 
the even number of quantum dots $N_{C}$.
%%%%%%%%
%odd-Nc%
%%%%%%%%
The situation is quite different for the array with odd $N_{C}$,
because the Kondo resonance contributes to the 
current as shown in Fig.\ \ref{fig:3dot}, where the series conductance
and  total charge $N_{\rm el}$ in triple dots ($N_C=3$) 
are plotted as functions of $\epsilon_d/U$.\cite{ONH} 
Near half-filling, the conductance shows a plateau of the height 
$g_{\rm series}^{\phantom{0}} \simeq 2e^2/h$ in stead of a wide gap $\Delta_{\rm gap}$ 
as seen in Fig.\ \ref{fig:U-dep}  for even $N_C$. 
The plateau becomes wider with increasing $U$.\cite{ONH} 
However, the Kondo temperature $T_K$ decreases with increasing $U$,\cite{OH}
and thus the conductance plateau can be observed only 
at low temperatures $T \lesssim T_K$. 
If the single-particle spectral function 
is calculated at half-filling as a function of the frequency $\omega$, 
the narrow Kondo peak will be seen at $0<|\omega| \lesssim T_K$.
Then, there will be a pseudo-gap 
region at $T_K\lesssim |\omega|<\Delta_{\rm gap}$, where 
the spectral weight almost vanishes. 
The width of the Kondo plateau at half-filling  
will be almost the same as  
 the excitation gap $2 \Delta_{\rm gap}$ seen in
the spectral function.
%%Large N
The Kondo temperature 
$T_K$ for half-filling decreases
with increasing  $N_C$, 
and finally the resonant state vanishes as  
$T_K\to 0$ in the limit of $N_C\to\infty$.
Therefore, for large $N_C$, 
the conductance plateau emerges only 
at very low temperatures.
If the thermodynamic limit $N_C\to \infty$ 
is taken first at finite temperature $T\neq 0$
and then the limit $T \to 0$ is taken next,
 the conductance will vanish because 
of the Mott-Hubbard behavior. 

%EOM
So far, we have concentrated on the transport properties 
at zero temperature. For comparison, we now consider the conductance 
at high temperatures. For qualitatively understanding of 
the physics at high-energy energy scale, 
the equation motion (EOM) method can be used.\cite{MWL}
It is equivalent basically to the Hubbard I approximation,\cite{Hubbard}
and in the present case we start with a {\it molecule limit}, 
where $\Gamma=0$ and 
the quantum-dot array can be regarded like an artificial {\it molecule\/}
described by the Hamiltonian ${\cal H}_C^0 +{\cal H}_C^U$ 
defined in eqs.\ (\ref{eq:HC^0}) and (\ref{eq:HC^U}).
With this method, the full Green's function 
 $G_{ij}^{({\rm EOM})}(\omega)$ is obtained 
 by substituting the self-energy defined with respect 
 to the {\it molecule\/}  $\Sigma_{ij}^{({\rm mol})}(\omega)$, 
which is calculated exactly, 
into the $N_C \times N_C$ matrix Dyson equation, 
\begin{equation}
\left\{ 
\mbox{\boldmath $G$}^{({\rm EOM})}(\omega)
\right \}^{-1}  = 
\left\{ 
\mbox{\boldmath $G$}^{(0)}(\omega)
\right \}^{-1}  -
\mbox{\boldmath $\Sigma$}^{({\rm mol})}(\omega) \;.
\end{equation}
Here, $\mbox{\boldmath $G$}^{(0)}(\omega) = \{G_{ij}^{(0)}(\omega)\}$ 
is the noninteracting Green's function 
defined with respect to the whole system including the two leads, 
and is determined by the Hamiltonian 
${\cal H}_{C}^0 +  {\cal H}_{\rm mix} + {\cal H}_{\rm lead}$. 
Then, a typical value of the conductance at high-temperatures 
 $T_K \ll T \ll U$ is estimated by using an 
 approximate formula following Kawabata,\cite{Kawabata}
\begin{equation}
g_{\rm series} \ \sim \  {e^2 \over h } \times   4\Gamma^2 
           \left| G_{N_C 1}^{({\rm EOM})}(0) \right| ^2 
\;,
\label{eq:EOM_g}
\end{equation}
where a factor $1/2$ has been introduced phenomenologically 
taking into account the fact that at high-energies 
the resonance tunneling using the Hubbard band occurs 
not simultaneously for the up and down spin components. 
In Fig.\ \ref{fig:EOM}, the conductance obtained from  
eq.\ (\ref{eq:EOM_g}) is plotted 
for (a) $N_C=3$ and (b) $N_C=4$, 
where 
the same parameter value is chosen for 
$U$ and $\Gamma$
 as those in 
Figs.\ \ref{fig:3dot} and \ref{fig:U1.0G0.12H0.1}.
At high temperatures, 
the conductance almost vanishes both for even and odd $N_{\rm el}$. 
The peaks of the Coulomb oscillation appear at the values of $\epsilon_d$, 
where $N_{\rm el}$ jumps discontinuously by an addition of one electron. 
If the temperature decreases, 
the bottom of the valleys at odd $N_{\rm el}$ will rise,
and it develops at $T \lesssim T_K$ 
to the Kondo plateaus of the Unitary limit 
which we have described in the previous section.
Particularly, the wide gap for the triple dots near half-filling seen 
in  Fig.\ \ref{fig:EOM} (a) evolves to the broad plateau in Fig.\ \ref{fig:3dot}.

%% the Hubbard bands for up and down spins 
%% do not contribute simultaneously to the resonance tunneling. 

%%%summary
In summary, we have studied the ground-state properties of 
an array consisting of four quantum dots based on  
a Hubbard chain attached to two non-interacting leads.
Using NRG approach, we have deduced the phase shifts, 
by which the conductance and local charge 
away from half-filling can be determined, 
from the low-energy eigenvalues near the Fermi-liquid fixed-point.
We have also discussed the parallel conductance 
of the quantum-dot array connected transversely to four leads.
Our formulation to calculate the two phase shifts 
for even and odd partial waves is quite general,
 and will be applied to various quantum-dot systems in the Kondo regime.

% 
%kouji
% The Mott-Hubbard behavior can be seen directly in the case of even $N_C$,
%  and it can be seen indirectly in the development 
%  of conductance plateau due to the Kondo resonance near the
% Fermi level, in the case of odd $N_C$.  

%
\begin{acknowledgments}
The authors are grateful to A.\ C.\ Hewson for valuable discussions.
Numerical computation was partly carried 
out using Computer Facility of  Yukawa Institute.
\end{acknowledgments}

%\appendix

%\section{Appendixes}

%% \bibliography{nrg-4v4}% Produces the bibliography via BibTeX.

\end{document}